\documentstyle[prl,aps,epsf,floats]{revtex}  

\newcommand{\nn}{\nonumber}

\begin{document}
\twocolumn[\hsize\textwidth\columnwidth\hsize\csname@twocolumnfalse\endcsname

\title{ Wigner symmetry in the limit of large scattering lengths 
}

\author{Thomas Mehen, Iain W.\ Stewart, and Mark B. Wise \\[4pt]}
\address{\tighten California Institute of Technology, Pasadena, CA, USA 91125 }

\maketitle

{\tighten
\begin{abstract}

We note that in the limit where the $NN$ $^1S_0$ and $^3S_1$ scattering lengths,
$a^{(^1S_0)}$ and $a^{(^3S_1)}$, go to infinity, the leading terms in the effective field
theory for strong $NN$ interactions are invariant under Wigner's SU(4) spin-isospin
symmetry.  This explains why the leading effects of radiation pions on the S-wave
$NN$ scattering amplitudes vanish as $a^{(^1S_0)}$ and $a^{(^3S_1)}$ go to infinity. 
The implications of Wigner symmetry for $NN \to NN\, \mbox{axion}$ and $\gamma\,
d\to n\, p$ are also considered.

\end{abstract}
}
\vspace{0.7in}
]\narrowtext

\newpage

Effective field theory methods are applicable to nuclear physics \cite{W1,W2}.  Recently
a new power counting has been developed for effective field theory in the two-nucleon
sector\cite{ksw,Bira}.   It is appropriate to the case where the scattering lengths
$a^{(^1S_0)}$ and $a^{(^3S_1)}$ in the $^1S_0$ and $^3S_1$ channels are large.  As
$a^{(^1S_0)}$ and $a^{(^3S_1)}$ go to infinity the couplings for the lowest dimension
2-body operators flow to a nontrivial fixed point\cite{W2,ksw}.  Higher dimension
2-body operators (and if the pion is not integrated out, pion exchange) are corrections
that can be treated perturbatively.  Neglecting these corrections the effective field
theory is scale invariant when the scattering lengths go to infinity.  In this paper we
note that in this limit the theory is also invariant under Wigner's SU(4) spin-isospin
transformations \cite{Wigner} 
\begin{eqnarray} 
  {\tighten \delta N = i  \alpha_{\mu\nu}\, \sigma^\mu \, \tau^\nu \, N \,, 
  \qquad N=\bigg(\begin{array}{c} p \\ n \end{array} \bigg)  \, .  } \label{trnfm} 
\end{eqnarray} 
In Eq.~(\ref{trnfm}), $\sigma^\mu = (1,\vec\sigma)$, $\tau^\nu=(1,\vec\tau)$, and
$\alpha_{\mu\nu}$ are infinitesimal group parameters (we will use the notation that
greek indices run over $\{0,1,2,3\}$, while roman indices run over $\{1,2,3\}$).  The
$\sigma$ matrices act on the spin degrees of freedom, and the $\tau$ matrices act on
the isospin degrees of freedom.  (Actually the transformations in Eq.~(\ref{trnfm})
correspond to the group SU(4)$\times$U(1). The additional U(1) is baryon number and
corresponds to the $\alpha_{00}$ term.)

Consider first the effective field theory for nucleon strong interactions with the pion
degrees of freedom integrated out.  The Lagrange density is composed of nucleon
fields and has the form ${\cal L}={\cal L}_1+{\cal L}_2 + \ldots$, where ${\cal L}_n$
denotes the n-body terms.  We have
\begin{eqnarray} \label{LN1}
  {\cal L}_1 &=&  N^\dagger \Big[ i\partial_t + {\overrightarrow\nabla^2/(2M) }
    \Big] N  + \ldots \,, \nn \\
  {\cal L}_2 &=& -\sum_{s}\,{C_0^{(s)} ( N^T P^{(s)}_i N)^\dagger ( N^T P^{(s)}_i N)} 
    + \ldots \,,
\end{eqnarray} 
where $M$ is the nucleon mass and the ellipses denote higher derivative terms.  
Here $s={^1S_0}\mbox{ or }{^3S_1}$, and the matrices $P_i^{(s)}$ project onto spin 
and isospin states
\begin{eqnarray}
  P_i^{({}^1\!S_0)} = { (i\sigma_2) \, (i\tau_2 \tau_i ) \over \sqrt{8} } \ ,\quad 
  P_i^{({}^3\!S_1)} = { (i\sigma_2 \sigma_i  ) \, (i\tau_2) \over \sqrt{8} }  \ .
\end{eqnarray}
The Lagrange density ${\cal L}_2$ can also be written in a different operator basis:
\begin{eqnarray}
  {\cal L}_2 &=& -\frac12 \Big[ {C_0^{S}}  ( N^\dagger  N )^2 
    + {C_0^{T}}( N^\dagger \vec \sigma N )^2 \Big] + \ldots \,,
\end{eqnarray}
where $C_0^{(^1S_0)}=C_0^S-3 C_0^T$ and $C_0^{(^3S_1)}=C_0^S+C_0^T$.  In this
basis it is the $C_0^T$ term that breaks the SU(4) symmetry (as well as some of the 
higher derivative terms).

Neglecting higher dimension operators in Eq.~(\ref{LN1}) the $^1S_0$ and $^3S_1$ 
$NN$ scattering amplitudes arise from the sum of bubble Feynman diagrams shown in
Fig.~\ref{C0b}.
\begin{figure}[!b]
  \centerline{\epsfxsize=14.5truecm \epsfbox{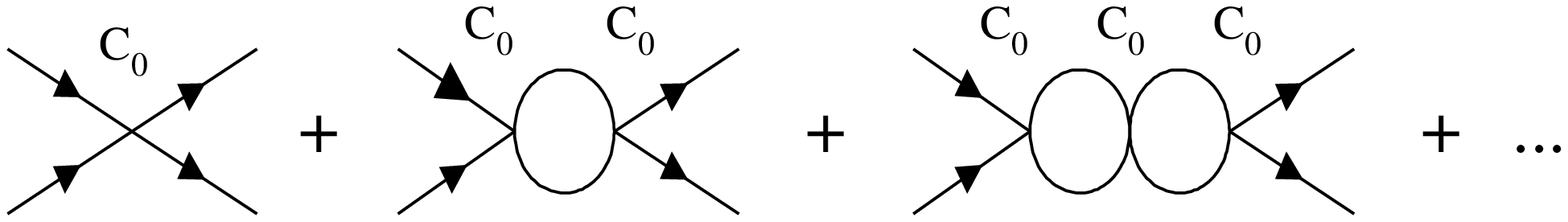}  }
 {\tighten
\caption[1]{The leading order contribution to the $NN$ scattering amplitude.} 
\label{C0b} }
\end{figure}
The loop integration associated with a bubble has a linear ultraviolet divergence and
consequently the values of the coefficients $C_0^{(s)}$ depend on the subtraction
scheme adopted.  In this paper we use dimensional regularization as the regulator.  In
minimal subtraction the coefficients are subtraction point independent and the center
of mass scattering amplitude is 
\begin{eqnarray} \label{AMS}
    {\cal A}^{(s)} = { -\bar{C_0}^{(s)} \over 1 + i \frac{Mp}{4\pi} \, \bar{C_0}^{(s)}  } \,,
\end{eqnarray}
where the bar is used to denote minimal subtraction and $p$ is the magnitude of the
nucleon momentum.  The S-wave scattering amplitudes can be expressed in terms of
the phase shifts $\delta^{(s)}$,
\begin{eqnarray}  \label{Apcotd}
    {\cal A}^{(s)} = \frac{4\pi}{M}\, \frac1{p\cot{\delta^{(s)}} -i p} \,,
\end{eqnarray}
and it is conventional to expand $p\cot{\delta^{(s)}}$ in a power series in $p^2$
\begin{eqnarray} \label{ere}
    p\cot{\delta^{(s)}} =- \frac{1}{a^{(s)}} + \frac12\, r_0^{(s)}\, p^2 + \ldots \,,
\end{eqnarray}
where $a^{(s)}$ is the scattering length and $r_0^{(s)}$ is the effective range\footnote{
\tighten Strictly speaking Eq.~(\ref{Apcotd}) only holds in the $^1S_0$
channel.  The $^3S_1$ channel is more complicated because of $^3S_1-^3D_1$
mixing, however the mixing is a small effect.}.  Comparing Eq.~(\ref{AMS}) with
Eqs.~(\ref{Apcotd}) and (\ref{ere}) we see that keeping only the lowest dimension two
body terms corresponds to neglecting the effective range and the higher powers of
$p^2$ in Eq.~(\ref{ere}), 
\begin{eqnarray}
   {\cal A}^{(s)} = -\frac{4\pi}{M}\, \frac1{ 1/a^{(s)} +i p} \,,
\end{eqnarray}
and that 
\begin{eqnarray}
  \bar{C_0}^{(s)} = \frac{4\pi a^{(s)}}{M} \,.
\end{eqnarray}
If $a^{(s)}$ is of natural size then the dimension six operators in Eq.~(\ref{LN1}) are
irrelevant operators.  It is then appropriate to perform a perturbative expansion of the
amplitude in a power series in $\bar{C_0}^{(s)}$, which corresponds to an expansion in
$p\, a^{(s)}$.  Terms cubic in $\bar{C_0}^{(s)}$ are not more important than the tree
level contribution of two body operators with two derivatives.  This situation would be
similar to the familiar application of chiral perturbation theory to $\pi\,\pi$ scattering. 
However, in nature the scattering lengths are very large: $a^{(^1S_0)}=-23.714 \pm
0.013 \,{\rm fm}$ and $a^{(^3S_1)}=5.425 \pm 0.001\, {\rm fm}$, or
$1/a^{(^1S_0)}=-8.3\,{\rm MeV}$ and $1/a^{(^3S_1)}= 36\,{\rm MeV}$ \cite{burcham}. 
The coefficients $\bar{C_0}^{(s)}$ are large and are very different in the $^1S_0$ and
$^3S_1$ channels.  Nonetheless, for $p\gg 1/a^{(s)}$ the amplitudes become, ${\cal
A}^{(s)}=4\pi i/(Mp)$.  The equality of the $^1S_0$ and $^3S_1$ amplitudes is
consistent with expectations based on Wigner symmetry.  The $p$-dependence is
consistent with expectations based on scale invariance\footnote{\tighten The scale
transformations appropriate for the non-relativistic theory are $x\to \lambda x$, $t\to
\lambda^2 t$, and $N \to \lambda^{-3/2} N$.}, since the cross section 
$\sigma^{(s)}=4\pi/p^2$.

In minimal subtraction, if $p\gg 1/a^{(s)}$ successive terms in the perturbative series
represented by Fig.~\ref{C0b} get larger and larger. Subtraction schemes have been
introduced where each diagram in Fig.~\ref{C0b} is of the same order as the sum.  It is
in these ``natural'' schemes that the fixed point structure of the theory and Wigner
spin-isospin symmetry are manifest in the Lagrangian.  One such scheme is PDS
\cite{ksw}, which subtracts not only poles at $D=4$, but also the poles at $D=3$ (which
correspond to linear divergences).  Another such scheme is the OS momentum
subtraction scheme \cite{W2,ms0}.  In these schemes the coefficients are subtraction
point dependent, $C_0^{(s)}\equiv C_0^{(s)}(\mu)$.  Calculating the bubble sum in PDS
or OS gives
\begin{eqnarray}
   {\cal A}^{(s)} = -{ C_0^{(s)}(\mu) \over 1 + \frac{M}{4\pi}(\mu+ip)C_0^{(s)}(\mu) }\,,
\end{eqnarray}
where
\begin{eqnarray}
   C_0^{(s)}(\mu) = -\frac{4\pi}{M} \frac1{\mu -1/a^{(s)}} \,.
\end{eqnarray}
For $\mu\sim p$ the contribution of every diagram in the sum in Fig.~\ref{C0b} is
roughly the same size.  Furthermore, as $a^{(s)}\to \infty$ the coefficients
$C_0^{(s)}(\mu)\to -4\pi/(M\mu)$ which is the same in both channels.  In this limit
$C_0^T(\mu)=[C_0^{(^3S_1)}(\mu)-C_0^{(^1S_0)}(\mu)]/4=0$ and
\begin{eqnarray}  \label{L3}
   {\cal L}_2 = -{2\pi \over M\mu} (N^\dagger N)^2 + \ldots \,.
\end{eqnarray}
The first term in Eq.~(\ref{L3}) is invariant under the Wigner spin-isospin
transformations in Eq.~(\ref{trnfm}).  The ellipses in Eq.~(\ref{L3}) denote terms with
derivatives, and they will not be invariant under Wigner symmetry even in the limit
$a^{(s)}\to \infty$.  However, these terms are corrections to the leading order Lagrange
density and their effects are suppressed by powers of $p/\Lambda$ (where $\Lambda$
is a scale determined by the pion mass and $\Lambda_{\rm QCD}$).  In the region
$1/a^{(s)} \ll p \ll \Lambda$ Wigner spin-isospin symmetry is a useful approximation
and deviations from this symmetry are suppressed by $C_0^T(\mu) \propto
(1/a^{(^1S_0)}-1/a^{(^3S_1)})$ and by powers of $p/\Lambda$. The measured effective
ranges are $r_0^{(^1S_0)}\!=\! 2.73\pm 0.03\,{\rm fm}$ and $r_0^{(^3S_1)}\!=\! 1.749\pm
0.008\,{\rm fm}$ \cite{burcham}.  A rough estimate of the scale is $1/\Lambda \sim
[r_0^{(^1S_0)} - r_0^{(^3S_1)}]/2 = 0.49\,{\rm fm}$, or $\Lambda\sim 400\,{\rm MeV}$. In
PDS or OS, the limit $a^{(s)} \to \infty$ is clearly a fixed point of $C_0^{(s)}(\mu)$ since
$\mu\, {\partial}/{\partial\mu} \, [\mu\, C_0^{(s)}(\mu)] =0$.  Also, scale invariance is
manifest since $\mu\to \mu/\lambda$ under scale transformations.  

Wigner symmetry is useful even though $a^{(^1S_0)}$ and $a^{(^3S_1)}$ are very
different.  This is because for $1/a^{(s)} \ll p \ll \Lambda$ corrections to the symmetry
limit go as $(1/a^{(^1S_0)}-1/a^{(^3S_1)})$ rather than $(a^{(^1S_0)}-a^{(^3S_1)})$. 
This is similar to the heavy quark spin-flavor symmetry of QCD \cite{HQS}, which
occurs in the $m_Q\to \infty$ limit. Heavy quark symmetry is a useful approximation for
charm and bottom quarks even though $m_b/m_c\simeq 3$.

As an application of the symmetry consider $NN\to NN\,\mbox{axion}$, which is
relevant for astrophysical bounds on the axion coupling \cite{axionguy}.  The axion is
essentially massless.  If the axion has momentum $\vec k$, and the initial nucleons
have momenta $\vec p$ and $-\vec p$ then the final state nucleons have momenta
$\vec q-\vec k/2$ and $-\vec q-\vec k/2$.  Energy conservation implies that
$p^2/M=q^2/M+k^2/(4M)+k$ where $p=|\vec p|, q=|\vec q|$, and $k=|\vec k|$.  In the
kinematic region we consider $q,p \gg k$, and the axion momentum can be neglected
in comparison with the nucleon momenta.  In this limit the terms in the Lagrange
density which couple the axion to nucleons take the form
\begin{eqnarray}
  {\cal L}_{int} &=& g_0  \Big(\nabla^j X^0\Big)\bigg|_{\vec x=0} \!\!\! N^\dagger 
  \sigma^j N +
   g_1  \Big(\nabla^j X^0 \Big)\bigg|_{\vec x=0}\!\!\!  N^\dagger \sigma^j\, \tau^3 N 
   \,,\nn\\
 & &
\end{eqnarray}

\newpage
\widetext \phantom{testtesttest}
\begin{figure}[!t]
  \centerline{\epsfysize=5.8truecm \epsfbox{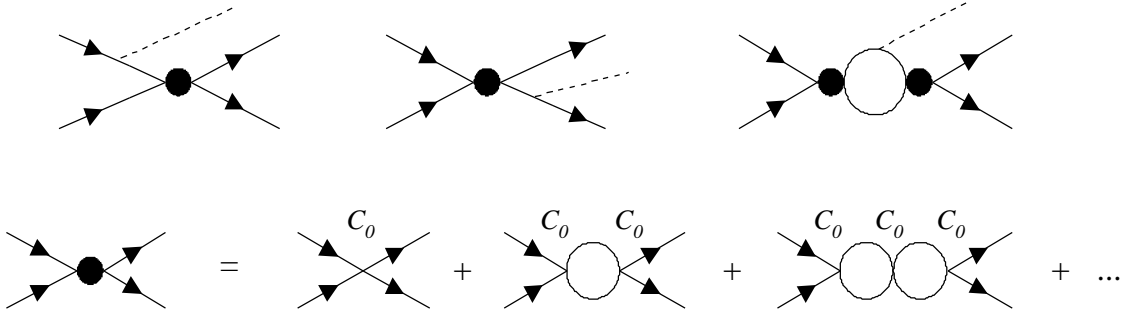}  }
 {\tighten
\caption[1]{Graphs contributing to $NN \to NN\,\mbox{axion}$ at leading order.  
The solid lines denote nucleons and the dashed lines are axions.} \label{axion} }
\end{figure}  \narrowtext
\noindent where $X^0$ is the axion field and $g_0,g_1$ are the axion-nucleon
isosinglet and isovector coupling constants.  Associated with spin-isospin symmetry
are the conserved charges
\begin{eqnarray}
   Q^{\mu\nu} = \int d^3x  N^\dagger \sigma^\mu \tau^\nu N \,,
\end{eqnarray}
and the axion terms in the action are proportional to these charges
\begin{eqnarray} \label{Saxion}
  S_{int} &=& g_0 \int\!\! dt \Big(\nabla^j X^0\Big)\bigg|_{\vec x=0}\!\! Q^{j0} +
    g_1 \int\!\! dt \Big(\nabla^j X^0\Big)\bigg|_{\vec x=0}\!\! Q^{j3} \,. \nn\\
 &&
\end{eqnarray}
The charge $Q^{j0}$ is the total spin of the nucleons which is conserved even without
taking the $a^{(s)}\to \infty$ limit, however $Q^{j3}$ is only conserved in the $a^{(s)}\to
\infty$ limit (and also in the limit $a^{(^1S_0)}\to a^{(^3S_1)}$).  Since conserved
charges are time independent, only a zero energy axion couples in Eq.~(\ref{Saxion}),
and these terms will not contribute to the scattering amplitude.  We conclude that
$NN(^1S_0)\to NN(^3S_1) X^0$ vanishes in the limit $a^{(s)}\to \infty$ and that
$NN(^3S_1)\to NN(^3S_1)X^0$ vanishes for all scattering lengths\footnote{\tighten
$NN(^1S_0)\to NN(^1S_0)X^0$ vanishes due to angular momentum conservation since
the axion is emitted in a P-wave.}.  Calculation of the Feynman diagrams in
Fig.~\ref{axion} shows that the leading order $^3S_1\to {^3S_1}$ scattering amplitude
does indeed vanish, and the $NN(^1S_0)\to NN(^3S_1) X^0$ amplitude is 
\begin{eqnarray}
  {\cal A} &=& g_1\, {4\pi \over M} \,  {\vec k \cdot \vec\epsilon\,^* \over k } 
  \bigg[ \frac1{a^{(^1S_0)}}-\frac1{a^{(^3S_1)}} \bigg] \bigg[{1 \over 
  1/{a^{(^1S_0)}}+i\,p}\bigg] \nn\\
  & &\qquad\qquad \times \bigg[{1 \over 1/{a^{(^3S_1)}}+i\,q} \bigg] \,,
\end{eqnarray}

\phantom{extraspacefiller} \vspace{2.80in}

\noindent where $\vec \epsilon$ is the polarization of the final $^3S_1$ $NN$ state. 
This is proportional to $(1/a^{(^1S_0)}-1/a^{(^3S_1)})$ and is consistent with our
expectations based on the Wigner symmetry.

Coupling of photons to nucleons occurs by gauging the strong effective field theory
and by adding terms involving the field strengths $\vec E$ and $\vec B$.  In the
kinematic regime where the photon's momentum is small compared to the
nucleons' momentum the part of the action involving the field strengths is
\begin{eqnarray} \label{SEB}
  S_{int} = {e\over 2M} \int dt\, {B^j}\Big|_{\vec x=0} \bigg( \kappa_0 Q^{j0}+
     \kappa_1 Q^{j3}  \bigg) + \ldots \,,
\end{eqnarray} 
where $\kappa_0$ and $\kappa_1$ are the isosinglet and isovector nucleon magnetic
moments in nuclear magnetons, and the ellipses denote subdominant terms.  The term
proportional to $\kappa_1$ in Eq.~(\ref{SEB}) gives the lowest order contribution to the
amplitude for $\gamma d \to n p(^1S_0)$.  The form of the coupling above implies that
like the axion case, this amplitude is proportional to $(1/a^{(^1S_0)}-1/a^{(^3S_1)})$.

So far we have considered an effective field theory with the pions integrated out.  It is
straightforward to include the pion fields and this is expected to increase the range of
validity of the momentum expansion.  Pion exchange can be separated into two types,
potential and radiation.  Potential pions have $k^0 \sim k^2/M$, where $k^0$ is the pion
energy and $k$ is the magnitude of the pion momentum.  Radiation pions are nearly
on-shell; i.e. $k^0 \sim \sqrt{k^2 + m_\pi^2}$. With the power counting in
Ref.~\cite{ksw}, $C_0^{(s)}(\mu)$ gives the leading order S-wave $NN$ scattering
amplitude, while potential pion exchange and four-nucleon operators with two
derivatives enter at next-to-leading order.  As our last example, we discuss the
corrections to $NN$ scattering due to radiation pions\cite{ms2}.  As pointed out in
Ref.\cite{ms2}, one should perform a multipole expansion on the coupling of radiation
pions to nucleons.  The first term in the multipole expansion is\footnote{Radiation
gluons in NRQCD and radiation photons in NRQED are also treated in this 
way\cite{gr}.}:
\begin{eqnarray}
  S_{int} = -{g_A \over \sqrt{2} f}  \int dt \,\Big(\nabla^i \pi^j\Big)\bigg|_{\vec x=0} Q^{ij}
    \,,
\end{eqnarray}
where $g_A\simeq 1.25$ is the axial coupling and $f\simeq 131\,{\rm MeV}$ is the pion
decay constant.  Radiation pions also couple to a conserved charge of the Wigner
symmetry in the large scattering length limit.  (A multipole expansion is not performed
on the coupling to potential pions so they do not couple to a conserved charge.) This
implies that only a radiation pion with $k^0=0$ will couple, which is incompatible with
the condition $k^0 \sim \sqrt{k^2 + m_\pi^2}$, so in the symmetry limit radiation pions
do not contribute to the scattering matrix element. In Ref.\cite{ms2}, it was shown by
explicit computation that graphs with one radiation pion and any number of
$C_0^{(s)}$'s give a contribution that is suppressed by at least one power of
$1/a^{(^3S_1)}-1/a^{(^1S_0)}$.  This suppression was the result of cancellations
between many different Feynman diagrams.  Wigner symmetry guarantees that the
leading contribution of graphs with an arbitrary number of radiation pions are
suppressed by inverse powers of the scattering lengths.

It has also been shown that Wigner symmetry is obtained in the large number of colors
limit of QCD \cite{ks}.  The implications of Wigner symmetry in nuclear physics were
studied in Ref.~\cite{WN}. So far the analysis in this paper has been specific to the
two-nucleon sector, however Wigner symmetry is observed in some nuclei with many
nucleons.  Terms with no derivatives also occur in ${\cal L}_3$ and ${\cal L}_4$, while
higher body contact interactions vanish because of Fermi statistics.  Fermi statistics
implies that there is only one four body term, $(N^\dagger N)^4$, which is invariant
under Wigner symmetry.  Furthermore, there is only one term in ${\cal L}_3$,
$(N^\dagger N)^3$, which is also invariant \cite{pc3}.  To see this, note that the three
nucleon and anti-nucleon fields must be combined in an antisymmetric way.  The three
$N$'s ($N^\dagger$'s) combine to a $\bar 4$ (4) of SU(4).  Combining the $4$ and $\bar
4$ gives $1 \oplus 15$, however only the singlet is invariant under the spin and isospin
SU(2) subgroups \cite{pc3}.  

Recent progress \cite{3bdy} in the three body sector suggests that the $(N^\dagger
N)^3$ contact interaction is not subleading compared with the effects of the first two
body term in Eq.~(\ref{L3}).  If the higher body operators with derivatives can be treated
as perturbations, then this letter shows that approximate Wigner symmetry in nuclear
physics is a consequence of the large $NN$ scattering lengths (and some simple group
theory).  

This work was supported in part by the Department of Energy under grant number
DE-FG03-92-ER 40701.  T.M. was also supported by a John A.  McCone
Fellowship.

{\tighten

} 


\begin{references}

\bibitem{W1} S. Weinberg, Phys.\ Lett.\ {\bf B251} (1990) 288.

\bibitem{W2} S. Weinberg, Nucl.\ Phys.\ {\bf B363} (1991) 3.

\bibitem{ksw} D.B. Kaplan, M.J. Savage, and M.B. Wise,  Phys. Lett. {\bf B424}
(1998) 390;  Nucl. Phys. {\bf B534} (1998) 329.

\bibitem{Bira} U. van Kolck, hep-ph/9711222 and Nucl. Phys. {\bf A645} (1999) 273.

\bibitem{Wigner} E. Wigner, Phys. Rev. {\bf 51} (1937) 106, 947;  {\it ibid.} {\bf 56}
(1939) 519. 

\bibitem{burcham} {\it Elements of Nuclear Physics}, W.E. Burcham, John Wiley \& Sons
Inc., (1979).

\bibitem{ms0} J. Gegelia, nucl-th/9802038; T. Mehen, and I.W. Stewart, Phys. Lett. 
{\bf B445} (1999) 378; T. Mehen and I. Stewart, nucl-th/9809095.

\bibitem{HQS} N.~Isgur and M.B.~Wise, Phys. Lett. {\bf B232} (1989) 113; Phys. Lett.
{\bf B237} (1990) 527.

\bibitem{axionguy} {\it Stars as Laboratories for Fundamental Physics}, G.G. Raffelt, 
The University of Chicago press, (1996). 

\bibitem{ms2} T. Mehen and I.W. Stewart, nucl-th/9901064.

\bibitem{gr} G.T. Bodwin, E. Braaten, and G.P. Lepage, Phys.\ Rev.\ {\bf D55} (1997)
1125;  P. Labelle, Phys. Rev. {\bf D58}, (1998) 093013; M. Luke and A. Manohar, Phys.\
Rev.\ {\bf D55} (1997) 4129;  B. Grinstein and I. Rothstein, Phys. Rev. {\bf D57} (1998)
78;  M. Luke and M.J. Savage, Phys.\ Rev.\ {\bf D57} (1998) 413.

\bibitem{ks} D.B. Kaplan and M.J. Savage, Phys. Lett. {\bf B365} (1996) 244;  
D.B.  Kaplan and A.V. Manohar, Phys. Rev. {\bf C56} (1997) 76.

\bibitem{WN}  {\it Group Symmetries in Nuclear Structure}, J.C. Parikh, Plenum press,
(1978);  T.W. Donnelly and G.E. Walker, Ann. of Phys. {\bf 60} (1970) 209; {\it
Theoretical Nuclear Physics and Subnuclear Physics}, by J.D. Walecka, Oxford
University Press (1995);  J.P. Elliott, in {\it Isospin in Nuclear Physics}, page 73, ed. D.H.
Wilkinson, North Holland Publishing, (1969);  P. Vogel, M.R. Zirnbauer, Phys. Rev. Lett.
{\bf 57} (1986) 3148; P. Vogel, W.E. Ormand, Phys. Rev. {\bf C47} (1993) 623;   
Y.V. Gaponov, N.B. Shulgina, and D.M. Vladimirov, Nucl. Phys. {\bf A391} (1982) 93.

\bibitem{pc3} P.F. Bedaque, H.W. Hammer, and U. van Kolck, private communication.

\bibitem{3bdy} P.F. Bedaque and U. van Kolck, Phys. Lett. {\bf B428} (1998) 221; P.F.
Bedaque, H.W. Hammer, and U. van Kolck, Phys. Rev. {\bf C58} (1998) R641, Phys. Rev.
Lett. {\bf 82} (1999) 463, Nucl. Phys. {\bf A646} (1999) 444.

\end{references}
\end{document}